`%\documentclass[prd,amsmath,amssymb]{revtex4}
\documentclass[aps,prl,twocolumn,showpacs,superscriptaddress,groupedaddress]{revtex4}  
\usepackage{graphicx}
\usepackage{dcolumn}
\usepackage{bm}
\usepackage{amssymb}   
\begin{document}

\title{Bounds on Spectral Dispersion from Fermi-detected Gamma Ray Bursts}

% [Submitted to: Physical Review Letters]

\author{Robert J. Nemiroff}
\author{Ryan Connolly}
\author{Justin Holmes}
\author{Alexander B. Kostinski}
\affiliation{Dept. of Physics, Michigan Technological University, 1400 Townsend Dr., Houghton MI, 49931, USA}

\begin{abstract}
Data from four Fermi-detected gamma-ray bursts (GRBs) is used to set limits on spectral dispersion of electromagnetic radiation across the universe.  The analysis focuses on photons recorded above 1 GeV for Fermi detected GRB 080916C, GRB 090510A, GRB 090902B, and GRB 090926A because these high-energy photons yield the tightest bounds on light dispersion.  It is shown that significant photon bunches in GRB 090510A, possibly classic GRB pulses, are remarkably brief, an order of magnitude shorter in duration than any previously claimed temporal feature in this energy range. Although conceivably a $>3 \sigma$ fluctuation, when taken at face value, these pulses lead to an order of magnitude tightening of prior limits on photon dispersion. Bound of $\Delta c / c < 6.94$ x $10^{-21}$ is thus obtained.  Given generic dispersion relations where the time delay is proportional to the photon energy to the first or second power, the most stringent limits on the dispersion strengths were $k_1 <$ 1.61 x $10^{-5}$ sec Gpc$^{-1}$ GeV$^{-1}$ and $k_2 <$ 3.57 x $10^{-7}$ sec Gpc$^{-1}$ GeV$^{-2}$ respectively.  Such limits constrain dispersive effects created, for example, by the spacetime foam of quantum gravity.  In the context of quantum gravity, our bounds set $M_1 c^2$  greater than 525 times the Planck mass, suggesting that spacetime is smooth at energies near and slightly above the Planck mass.
\end{abstract}
\maketitle

% Main Body of Paper [PRL does not have sections]

Gamma-ray bursts (GRBs) are the furthest known explosions in the universe.   Their rapid variability and great distances make them useful as probes of light properties as well as the intervening space.  Were light to have fundamentally different speeds at different wavelengths (spectral dispersion), distant GRBs might show persistent energy-dependent arrival patterns \cite{Gam99}.  Spacetime foam inherent in some formulations of quantum gravity, for example, might cause spectral dispersion \cite{Mav11, Ame10, Kos08}.  Other properties of light or the universe might also cause different wavelengths to propagate at different speeds \cite{Bie09, Ellis00}.

GRBs have already been used to limit the cosmological density of compact objects through the non-detection of their gravitational lensing \cite{Nem01}.  Lag-minimizing algorithms have been previously designed to search for quantum-gravity based dispersion effects \cite{Sca08}.  Although bounds on quantum gravity dispersion in Fermi GRBs have been explored previously for two different Fermi GRBs \cite{Abd09B, Abd09}, the present work limits more general parameters, considers four Fermi GRBs, considers only super-GeV photons, and yields substantially tighter bounds.

Given that two photons of different energies $\Delta E$ are emitted at the same place and time, the gap $\Delta t$ between their arrivals can be quantified as
 \begin{equation}
 \Delta t = k_n D_n E^{n-1} \Delta E ,
 \end{equation}
where $k_n$ is the dispersion strength and $D_n$ is a cosmological lookback distance that also depends on the nature of the photon dispersion \cite{Jac08}.  Specifically,
\begin{equation}
D_n = {c \over H_o} \int_0^z { (1 + z')^n \ dz' \over
                    \sqrt{ \Omega_M (1 + z')^3 + \Omega_{\Lambda} } } ,
\end{equation}
where $H_o$ is present value of Hubble's constant, and $\Omega_M$ and $\Omega_{\Lambda}$  are the present values of the matter density and cosmological constant density \cite{Jac08, Hog99} in a geometrically flat universe \cite{Nem08}.

For clarity and following  theoretical precedents \cite{Sch99, Ellis08, Jac08}, only three cases will be considered here: $n = -1$, $n = 1$ and $n = 2$. The first case, $n = -1$, is for a universe with no chromatic dispersion.  Then, $k_{-1}=0$ and $D_{-1}$ corresponds with the classic cosmological lookback distance \cite{Hog99}.  In the second case, $n = 1$, the dispersion delay scales with the energy difference between photons, a primary case expected were spacetime to have the foaminess inherent in some models of quantum gravity \cite{Ellis08}.  The third case has $n = 2$, is considered in some models of quantum gravity \cite{Ellis08}.  It will be assumed here that dispersion occurs uniformly along the light paths.

For a group of photons emitted over a source of finite size, an upper limit on $\Delta t$ might relate primarily to an upper limit on source size and not to dispersion properties of light.  Given limited information, one might not be able to disentangle the various contributions to $\Delta t$.  Surely, though, an observed bound on $\Delta t$ would constrain the combined processes, thereby limiting the individual magnitudes.  An exception to this would be if the source and universe dispersion effects were of similar magnitudes but of opposite sign, a coincidence that is testable with a larger data set but here considered unlikely.

Because the largest energy ranges occur most commonly in the GRBs with the highest energy photons, and since these GRBs with many high energy photons are rare, GRBs with numerous high energy photons were initially sought -- to find the finest temporal feature of statistical significance.  A useful previous search included one by Rubtsov et al. \cite{Rub11} of the Fermi LAT photon database, although other previous studies also were influential \cite{Max11, Cor10, Abd09B}.   Another clue came from a visual inspection of Figure 1 of Abdo et al. (2009) \cite{Abd09}, where a striking clustering of photons above 1 GeV was spotted for GRB 090510A.  Other reasons for our 1 GeV threshold include the lower photon background at higher energies, and the possibility of extremely brief GRB pulses at higher energies.  Four candidate GRBs eventually emerged: GRB 080916C, GRB 090510A, GRB 090902B, and GRB 090926A.  ``Pass 7" data from these GRBs were downloaded from the Fermi web interface at NASA's GSFC in February 2012.  Only photons within a 95 \% energy-dependent error radius of the sky position of the optical counterpart were considered.  This error radius was interpolated from Fermi performance data given by Ref. \cite{LATPerf}.  
 
\begin{figure}
\includegraphics[scale=0.7]{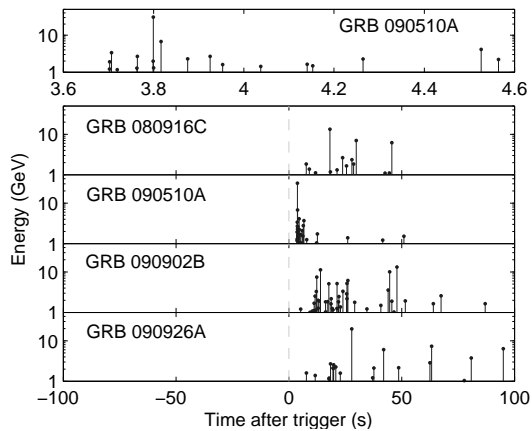}
\caption{\label{fig:fig1} Time series of photon arrivals for the four Fermi GRBs analyzed.  Top panel is a closeup of the one second of GRB 0901510A, containing the finest temporal features. }
\end{figure}

The bottom four panels of Fig. (1) show time series for the arrivals of photons of the four Fermi GRBs. The origin $t = 0$ indicates the time that the GRB triggered on Fermi's GLAST Burst Monitor (GBM).  On the left, at negative times, is 100 seconds of Fermi LAT data that occurred before the trigger time while on the right, at positive times, is 100 seconds of data that occurred after the trigger time. Individual counts are shown as vertical line segments.  The height of the line segment indicates the recorded energy of the photon detected. Inspection of Figure 1 shows that the background for stray photons, prior to the trigger time, for example, is very low.  The top panel of Fig. (1) shows a closeup of the one second of GRB 0901510A when the bunched photons arrived.

\begin{table*}
\caption{\label{tab:table2} Measured parameters of selected high energy Fermi GRBs.  }
\begin{ruledtabular}
\begin{tabular}{lcccccc}
GRB & $\Delta t$ & N & E (low) & E (high) & $z$ [a][Ref] \\
Name & sec  &  & GeV (2$\sigma$) & GeV (2$\sigma$) &  (2$\sigma$) \\
\hline
080916C & 37.9 & 14 & 1.32 & 10.6 & 4.05 \cite{Gre09}  \\
090510A & 0.00155 & 11 & 1.58 & 24.7 & 0.897 \cite{Abd09}  \\
090902B & 23.9 & 33 & 1.20 & 9.02 & 1.82 \cite{Cuc09}  \\
090926A & 3.00 & 7 & 1.38 & 2.15 & 2.106 \cite{Mal09}  \\
\end{tabular}
\end{ruledtabular}
\end{table*}

For a (short) GRB 090510A, consider the first 11 photons arriving over a $\Delta T = 0.1745$ seconds. The post-trigger arrival times of these photons were 3.702234, 3.702783, 3.706941, 3.719431, 3.763108, 3.764177, 3.799190, 3.799318, 3.800096, 3.816729, 3.875767 sec, respectively. For comparison, the next five photons, photons 12 through 16, arrived at 3.925311, 3.953093,  4.037660, 4.140611, and 4.152783 sec.  The sixth photon had the unusually high energy of 30.9 GeV.  Of the 11 photons considered, 6 photons arrived before the temporal midpoint and 5 photons arrived later.  Notable is the closeness in arrival times of three photon groups. These groups are defined by the first and second photons, the fifth and sixth photons, and photons seven through nine.  The time between the first and last photons in these groups are 0.549 ms, 1.069 ms, and 0.906 ms respectively.

\begin{table*}
\caption{\label{tab:table3} Derived and limited parameters of selected high energy Fermi GRBs.  }
\begin{ruledtabular}
\begin{tabular}{lcccccccc}
GRB & $D_{-1}$ & $\Delta c / c$ & $D_1$ & $k_1$ & $M_1 c^2$ & $D_2$ & $k_2$ & $M_2 c^2$ \\
Name & Gpc & & Gpc & sec$/$(Gpc GeV) & GeV & Gpc & sec$/$(Gpc GeV$^2$) & GeV \\
\hline
080916C & 3.57 & 1.03E-16 & 16.8 & 2.42E-01 & 4.25E+17 & 48.8 & 6.28E-03 & 4.96E+09  \\
090510A & 2.17 & 6.94E-21 & 4.18 & 1.61E-05 & 6.41E+21 & 6.09 & 3.57E-07  & 6.57E+11 \\
090902B & 2.96 & 7.84E-17 & 8.38 & 3.65E-01 & 2.82E+17 & 16.0 & 1.70E-02  & 3.01E+09 \\
090926A & 3.10 & 9.40E-18 & 9.59 & 4.05E-01& 2.54E+17 & 19.5 & 7.41E-02 & 1.44E+09  \\
\end{tabular}
\end{ruledtabular}
\end{table*}

Is this arrival pattern of remarkably brief doublets separated by long pauses significant? Do these 3 brief pulses define the finest time scale yet?  We argue that such ``rhythm" is, most likely, not spurious. As shown below, this group of 11 photons is consistent with a constant overall arrival rate.  Yet, the following simple, albeit crude, analytical argument shows that the odds of a uniformly emitting source producing the pattern described above, are below 3 $\sigma$.  This is then confirmed by a detailed Monte Carlo simulation.

For a perfectly random (Poisson) process, the waiting times ($t$) between consecutive photon arrivals are exponentially distributed and a sum of $m$ such times is $\Gamma$-distributed with exponent $m$ (convolution of $m$ exponential variates).  Given an estimated mean waiting time $\tau = 0.1745/10$ sec, the probability of waiting $t << \tau$ is $t/\tau$.  For example, consider $t < 1.069$ ms a ``success".  The probability of success is then $\approx 0.1069 /1.750 = 0.0613$ for the 11 photon group.  Then the (binomial) probability of at least 4 ``successes" in 10 trials (10 waiting times between the 11 photons) is $P(4, 11) = 1/455$. If one counts the triplet as 3 successes, the odds drop to $P(5, 11) = 1/5000$. These crude estimates bracket the result of the $10^9$ uniformly random Monte Carlo runs, indicating that the chance that 5 photons would trail other photons by 1.069 ms or less occurs in only about 1 in 1190 trials (about 3.34 $\sigma$).  A sceptic might object that the mean rate need not be uniform, that both the 1st and the 11th arrivals ought to be regarded as fixed, etc.  To that end, we now describe our data analysis as well as more elaborate Monte Carlo simulations in more detail.

To determine the briefest yet statistically meaningful time interval $\Delta t$ in the data, we proceeded as follows.  Groups of consecutive photon arrival times were considered, starting from the three photons arriving closest in time, then the four closest photons, and subsequently all numbers of GRB-associated photons for 500 seconds following the trigger.  To ensure relatively uniform average arrival rates, we chose photon groups with roughly equal numbers of photon arrivals before and after the temporal midpoint of the group.  Formally, a two-bin $\chi^2$ statistic was computed. Given the single degree of freedom, ``flat" groups with $\chi^2 < 1$ were considered as statistically consistent with a flat distribution, and then search for $\Delta t$ proper ensued, aided by a Monte Carlo simulation as follows.

For each photon in the time series except the last, the number of trailing photons, arriving within a time window $T$ was counted, for a wide range of $T$s.  This photon count was compared to that expected from a uniformly random arrival time distribution.  The comparison distribution typically involved $10^6$ trial time-series.  To avoid spurious bunching, only  $\Delta t$ delays such that the associated number of real photons was found in less than 1 \% of the equivalent Monte Carlo distributions, were considered for further analysis.

Returning to the GRB 090510A 11 photon group, do the 3 brief pulses define the finest time scale of significance?  Could a variable mean rate, perhaps, produce such a pattern? To that end, we assumed that each of the 3 photon groups was randomly chosen from a single parent pulse form.  This pulse form is the generic GRB ``Norris pulse" shape first suggested by Norris \cite{Nor05} for the instance found most common by Nemiroff \cite{Nem11}, specifically, $P = A e^{-t/\alpha - \alpha/t}$ where $P$ is the photon count rate, $A$ is the pulse amplitude, $t$ is time during the pulse, and $\alpha$ is the time scale of the pulse.  To be conservative, we will focus on the broadest photon group, the central pair separated by 1.069 ms.

A simple simulation shows that randomly chosen pairs of photons from a Norris pulse form have a mean pair separation of about 1.20 $\alpha$.  Additionally, in a Norris pulse, 68.2 \% of the photons arrive within a total time window of 1.74 $\alpha$ surrounding the pulse peak, here called the pulse ``width".  Therefore, a parent pulse with width of $\Delta t = (1.74/1.20) 1.069$ ms $=$ 1.55 ms would yield a mean pair separation of 1.069 ms, the longest time between first and last photons of the three close photon groups of GRB 090510A.  Therefore, in subsequent analysis, we will use $\Delta t$ = 1.55 ms.

The conservative value of $\Delta t$ estimated above for GRB 090510A is about a factor of ten smaller than even the least conservative limit on $\Delta t$ listed by Ref. \cite{Abd09} in row 5 of Table S1.  A primary reason for this is that Ref. \cite{Abd09} measured the limiting $\Delta t$ essentially as the time difference between the start of a sub-MeV spike and a possibly associated 0.75 GeV photon.  Our analysis differs
from this earlier analysis of GRB 090510A in that they looked at photons over a wide range of energies, whereas we looked at only the most energetic photons ($>$1 GeV) because the pulse durations are known to decrease greatly as photon energy is increased, so the tightest limits on the dispersion delays will come from the highest energy photons.  Therefore, the small $\Delta t$ values presented here focus on extremely short doublets prominent at very high energies.

Of the four GRBs considered, only GRB 090510A and GRB 0900902B have photons arriving close enough in time to eclipse the 0.01 sec previously reported \cite{Abd09} as the smallest $\Delta t$ record.  We therefore conclude that analyzing the other GRBs at most increased the number of trials to two, which would decrease the statistical significance of the $\Delta t$ reported here for GRB 090510A to about 3.14 $\sigma$, still above 3 $\sigma$.

For GRB 08016C, GRB 090902B, and GRB 090926A, none of the photon groups for which the 2-bin $\chi^2$ test was less than unity showed significant bunching on any time scale.  On longer time scales, clearly distinct photon groups have their $\Delta t$ values recorded in Table 1.

Table 1 lists the measured parameters for the four GRBs selected.  Column 1 lists the title of the GRB, coded with its date of detection.  The $\Delta t$ values as well as the number of photons $N$ on which they are based as listed in Columns 2 and 3 respectively.

Another measured parameter that limits spectral dispersion is $\Delta E$, the energy between the highest and lowest energy photons arriving from the GRB in the $\Delta t$ time window. Conservative 2 $\sigma$ values of the lowest and highest energy photons -- $E$ (low) and $E$ (high) -- are given, assuming a 10 \% single $\sigma$ energy measurement uncertainty.  They are listed in Columns 4 and 5 of Table 1.  Values for the GRB redshifts were obtained by others from follow-up observations of the GRB optical afterglows and the $2\sigma$ lower limits are listed in Column 6 of Table 1, with references.

The ratio of $\Delta t$ and $\Delta E$ has been used to set limits on Lorentz invariance previously, where Boggs et al.  \cite{Bog04} derived an upper limit of $\Delta t / \Delta E$ of 0.7 sec / GeV for GRB 021206.  For GRB 090510A, Ref. \cite{Abd09} list $\Delta t / \Delta E < 0.03$ sec / GeV at the 99 \% confidence level as their conservative limit (no least conservative limit is listed).  The tightest bound from Table 1, however, involving the upper limit on $\Delta t$ for GRB 090510A, is $\Delta t / \Delta E <$ 6.71 x 10$^{-5}$ sec / GeV, an improvement of greater than two orders of magnitude.

From the measured parameters listed in Table 1, derived and limited parameters were computed and listed in Table 2.  Values of $D_{-1}$, $D_1$, and $D_2$ were computed from Eq. (2) under the assumption of a flat concordance cosmology with $\Omega_M = 0.3$, $\Omega_{\Lambda} = 0.7$ and a Hubble constant $H_o$ of 72 km sec$^{-1}$ Mpc$^{-1}$, and are listed in Columns 2, 4, and 7 of Table 2, respectively.

Given the above data, it is possible to place bounds for the difference between the speeds of light at different energies: $\Delta c / c$.  Assuming $\Delta c$ results from an inherent property of electromagnetic radiation itself, then the lookback distance each photon has traveled is $D_{-1}$ as given by Eq. (2) \cite{Hog99}.  Defining lookback time as $t = D_{-1}/c$, the time differential yields $\Delta c / c = c \Delta t / D_{-1}$.  Limits on $\Delta c / c$, computed using our strictest upper limit on $\Delta t$, are listed in Column 3 of Table 2.

A previous limit on $\Delta c / c$ using GRBs was obtained in 1999 by Schaefer \cite{Sch99}, where an analysis of GRB 930229 yielded $\Delta c / c <$ 6.3 x 10$^{-21}$ for photons of energies between 30 and 200 KeV.  A comparable limit for $\Delta c / c <$ 6.94 x 10$^{-21}$ is derived here from the $\Delta t$ listed in Column 2 of Table 1 for GRB 090510A for photons of energy difference $\Delta E \gtrsim$ 23.5 GeV.

Alternatively, it can be assumed that it is the intervening space that causes differential speed for photons of different energies.  Following Eq. (2) and approximating $E \sim \Delta E$, it is clear that $k_n < \Delta t / (D_n \Delta E^n)$.  In other words, were $k_n$ greater than this, the universe would have separated photons of an energy difference greater than $\Delta E$ by more than $\Delta t$.  For $n = 1$ and $n=2$, using the $\Delta t$ limits listed in Column 2 of Table 1,limiting $k_1$ and $k_2$ values are listed in Table 2's Columns 5 and 8 respectively.

The $k_1$ parameter effectively limits dispersion expected in some versions of quantum gravity \cite{Ellis08}.  In particular, given that $\Delta t \sim (\Delta E / M_1 c^2)(D_1/c)$ as delineated in Ref. \cite{Abd09}, then $M_1 c^2 = (k_1 c)^{-1}$.  In this parametrization, $M_1 c^2$ is a minimum energy scale of the inherent foaminess of spacetime responsible for the dispersion.  Note that the above data places an upper limit on $k_1$ which translates into a lower limit on $M_1 c^2$. Similarly, it is found that $M_2 c^2 = (3 k_2 c / 2)^{-1/2}$.  The limiting values of $M_1 c^2$ and $M_2 c^2$ are listed in Table 2's Columns 6 and 9 respectively.

Prior to Fermi, GRB  published lower limits for $M_1 / M_{Planck}$ and $M_2 / M_{Planck}$ were on the order of $0.04$ and 4 x 10$^{-12}$ respectively \cite{Bog04, Sch99},  where $M_{Planck} c^2 = 1.22$ x $10^{19}$ GeV.  Using Fermi data for GRB 090510A, however, Ref. \cite{Abd09} found $M_1/M_{Planck} > 102$, while this was relaxed to $M_1 / M_{Planck} > 1.19$ for more conservative assumptions.  Note that  using the most stringent upper limit on $\Delta t$ listed found here for conservative assumptions results in a rather tight bound of $M_1/M_{Planck} > 525$, suggesting that space is smooth even at energies near and slightly above the Planck mass.

RJN acknowledges helpful conversations with Jerry T. Bonnell and Jay P. Norris.  This work was supported, in part, by  NSF AGS - 31 1119164

\end{document}